\documentclass[preprint,12pt]{article}
\usepackage{amsfonts,epsfig}
\usepackage{amsmath,amsfonts}
\usepackage{amssymb,graphicx}

\def\pa{\partial}

\def\k{\kappa}
\def\d{\delta}
\def\h{\hat}
\def\b{\bar}
\def\t{\tilde}
\def\f{\frac}

\def\p{\varphi}
\def\k{\kappa}
\def\d{\delta}
\def\h{\hat}
\def\t{\tilde}
\def\f{\frac}

\def\l{\label}
\def\e{\varepsilon}

\def\la{\lambda}
\def\g{\gamma}

\def\m{\mu}
\def\n{\nu}

\def\r{\rho}
\def\s{\sigma}

\def\th{\theta}
\def\o{\omega}

\def\be{\begin{equation}}
\def\ee{\end{equation}}
\def\ba{\begin{eqnarray}}
\def\ea{\end{eqnarray}}
\def\z{\bar{z}}

\def\r{\rho}
\def\s{\sigma}

\def\th{\theta}
\def\o{\omega}

\def\be{\begin{equation}}
\def\ee{\end{equation}}
\def\ba{\begin{eqnarray}}
\def\ea{\end{eqnarray}}
\def\z{\bar{z}}
\def\c{\chi}

\begin{document}

\vspace{5cm}
\centerline{\large\bf Horizon symmetries of
 black holes with supertranslation field}
\vspace{1cm}
\centerline{{\bf Mikhail Z. Iofa}
\footnote {e-mail iofa@theory.sinp.msu.ru} }
\centerline{Skobeltsyn Institute of Nuclear Physics}
\centerline{Lomonosov Moscow State University}
\centerline{Moscow, 119991, Russia}
\vspace{1cm}

\begin{abstract}
Near-horizon symmetries are studied for black hole solutions to Einstein
equations containing  supertranslation field constructed
by Compere and Long. The metric is transformed to variables in which
the horizon is located at the surface $r=2M$, where $M$ is the mass of 
black hole.
We consider general  diffeomorphisms which preserve the gauge and the near-horizon
structure of the metric and find the corresponding transformations of 
 metric components.
We review the action of the generators of supertranslations 
preserving the static gauge of
the metric and determine a subgroup of supertranslations preserving 
the gauge and near-horizon structure of the metric.
 Variation of the surface charge corresponding to the
 Killing vectors of asymptotic horizon symmetries is calculated.
Sufficient conditions of integrability of the variation of the surface 
charge to a closed
integral form are found and an example of metrics with integrable
 charge variation is discussed. 
\end{abstract}
\vspace{3cm}

\section{Introduction}


BMS symmetry is a symmetry of the asymptotically flat spaces
 at the null infinity. The infinite-dimensional group of the BMS
transformations
extends the Poincare group and contains  supertranslations,
angular-dependent shifts of retarded time at null infinity
 \cite{bondi,sachs,strom1}.
 Finite supertranslation
 diffeomorphisms map field configurations to inequivalent, physically different
configurations which differ by structure of a cloud of soft particles
\cite{fad,carney,strom4}.
Configurations with different supertranslation fields are physically different in a
sense that their corresponding (superrotation) charges are different
\cite{bar3,bar2,bar4,charge,comp1,comp2}.

The final  state of gravitational collapse is diffeomorphic to the Kerr
space-time. The set of diffeomorphisms contains supertranslations, and the
resulting stationary metric contains  supertranslation field.
The final state of collapse is parametrisied
by mass, angular momentum and supertranslation field \cite{comp2}.

The BMS transformations are naturally formulated at the null infinity, and there is
a complicated problem of extension  of an asymptotically defined metric with
supertranslation field
in a closed form  to the bulk.
In paper \cite{comp1} a family of
vacua containing  supertranslation field was constructed in the bulk. In paper
\cite{comp2} a solution-generation technique was developed and
applied to construction of the exact solutions to Einsteins equations which are
black hole metrics containing supertranslation field.
It is an interesting question, how the infinite-dimensional symmetries 
of the space-time reveal themselves in the near-horizon physics.
   
Because many problems of black-hole physics are
connected with horizon structure of black holes, and having an explicit
example of a metric with supertranslation field,
in the present paper we study the near-horizon symmetries of the black hole metric
\cite{comp2}
containing supertranslation field. 

In Sect.2 the metric of \cite{comp2} is transformed to a form in which horizon
of the black hole is located at the surface $r=2M$, where $M$ is the mass 
of black hole.
Solving the geodesic equations, we show that the surface $r=2M$ 
is the surface of infinite red-shift.

In Sect.3 we study general near-horizon symmetries of the metric
preserving the form of the metric components in
the leading orders in distance from the horizon $r-2M$.
Solving the equations for the asymptotic Killing vectors
corresponding to  near-horizon symmetries of the metric,
we find variations of the metric components under the action
of generators of the asymptotic near-horizon symmetries.

In Sect.4, first, we review the
action of supertranslations preserving the static gauge of the metric.
Restricted to the near-horizon region, supertranslations act as near-horizon
symmetries. We find a subgroup of the group of supertranslations which preserve
both the gauge and the near-horizon form of the metric.

In Sect.5, using the results of Sect.3, 
we calculate  variation of the surface charge corresponding to the asymptotic
horizon symmetries. 
Sufficient conditions on the metric making possible integration of the variation
of the charge  over the space of metrics  to a closed form are found. 
Example of metrics with supertranslation field  depending
only on spherical angle $\th$ is considered.
  It is shown that in this case 
there appear relations between the
components of the metric which make possible  integration of the variation of the 
charge to a closed expression.

Sect.6 contains  remarks on connection of the results of the present
with other papers.

\section{Static metric with supertranslation field}

In this section, after a short review of black hole with supertranslation 
field constructed in \cite{comp2}, we transform the metric to a form with the horizon
located at the surface $r=2M$. Solving the geodesic equations for null geodesics
we show that the surface $r=2M$ is the surface of infinite red-shift.

Vacuum solution of the Einstein equations containing supertranslation field 
$C(z, \z )$ is
\ba
\l{2.1}
&{}& ds^2 =\t{g}_{mn}dx^m dx^n = \\\nonumber
&{}&= -\f{(1-M/2\r_s )^2}{(1+M/2\r_s )^2}dt^2 +
(1+M/2\r_s )^4 \left[d\r^2 + (((\r -E)^2 +U)\g_{ab} +(\r -E)C_{ab})dz^a dz^b
 \right],
\ea
where $z^a =z,\,\z$.
Supertranslation field  $C(z,\z )$ is 
a real regular function on the unit sphere.
Here
\be
\l{2.2}
\r_s (\r, C) =\sqrt{(\r -C -C_{00})^2 + D_a C D^a C}
.\ee
$C_{00}$ is the lowest spherical harmonic mode of $C(z,\z )$.
In the following we do not write $C_{00}$ explicitly understanding
$C\rightarrow C-C_{00}$.
Covariant derivatives $D_a ,\,\, a=z,\z $ are  defined
with respect to the metric on the sphere
 $ds^2 =\g_{z\b{z}}dz d\b{z}, \,\, z = \cot\f{\th}{2} e^{i\p},\,\,\,
\g_{z\b{z}}= 2e^{-2\psi},\,\, \psi =
\ln(1+|z|^2 )$. Also we consider the metric (\ref{2.1}) written in
spherical coordinates with the metric on the sphere  $ds^2 =d\th^2 +
\sin^2\th d\p^2$.
The tensor $C_{ab}$  and the functions
$U$ and $E$ are defined as
\ba
\l{2.3}
\nonumber
C_{ab} = -(2D_a D_b -\g_{ab} D^2 ) C,\\
U=\f{1}{8} C_{ab}C^{ba},\\\nonumber
E=\f{1}{2}D^2 C + C,
\ea
and in $z,\,\z$ representation are equal to
\ba
\l{2.4}
C_{zz} =-2D_z D_z C, \qquad C_{\z\z} =-2{D}_{\z} {D}_{\z} C, \qquad C_{z\z}= 0,\\
U= \f{1}{4}(\g^{z\z})^2 D^2_z C D^2_{\z}  C,\\
E =\g^{z\z}D_z D_{\z}  C  +C.
\ea
We introduce a new variable $r=r(\r , z^a ) $ chosen so that the  $\t{g}_{tt}$
component of the
metric is equal to $1-2M/r$ \cite{iofa}. 
Variable $r$ is defined the by the relation
\be
\l{2.5}
r=\r_s (\r, C)\left(1+\f{M}{2\r_s (\r, C)}\right)^2 
.\ee
Inversely,  $\r$ is expressed through $r$ as
\ba
\l{2.r}\nonumber
 \r=C +\sqrt{\f{K^2}{4}-D_a C D^a C},
\ea
where we introduced the functions
\be
\l{2.6}
K=r-M + r V^{1/2},\qquad\,\, V=1-\f{2M}{r}
.\ee
Introducing
\be
\l{2.9}
b_a = \f{2\pa_a C}{K} , \qquad b^2 = b_a b^a,\qquad a=z,\,\z
,\ee
we have the expression for $d\r (r, z^a )$ in a form
\ba
\l{2.12}
&{}&d\r =\f{K}{2\sqrt{1-b^2} }\left[ 
f_a dz^a
+ \f{dr}{rV^{1/2}}\right],\\
&{}& f_a =b_a\sqrt{1-b^2 } -\f{\pa_a b^2}{2}
.\ea
Writing
\ba
\l{2.13}
((\r -E)^2 +U)\g_{z\z}=
 \left(\f{K}{2}\right)^2 \h{g}_{z\z},\\
\l{2.14}
(\r - E)C_{zz}=
 \left(\f{K}{2}\right)^2\h{g}_{zz},\\
\l{2.15}
(\r - E)C_{\z\z}=
 \left(\f{K}{2}\right)^2\h{g}_{\z\z}.
\ea
we express the metric (\ref{2.1}) in a form
\ba
\nonumber
&{}&ds^2=g_{tt}dt^2 + g_{rr}dr^2+ g_{r z}dr dz +g_{r \z}dr d\z +g_{zz}dz dz
+g_{\z\z}d\z d\z +2g_{z\z}dz d\z=\\\nonumber
&{}&= -Vdt^2 + 
\f{dr^2}{ V(1-b^2 )}+\f{r (f_z dz +f_{\z} d\z ) dr}{V^{1/2}(1-b^2 )}
+\\\l{2.16}
&{}&+r^2\left[\left( \f{f^2_z }{1-b^2 } + \h{g}_{zz}\right )dz^2 +
\left( \f{f^2_{\z}}{1-b^2 } + \h{g}_{\z\z }\right )d{\z}^2 
+2\left(\f{f_z f_{\z}}{1-b^2 } +\h{g}_{z \z}\right )dz d{\z}\right]
.\ea
To obtain the metric in a form (\ref{2.16}),  we have used the relations
\be
\l{2.17}
\f{(1-M/2\r_s )^2}{(1+M/2\r_s )^2}=V, \qquad (1+M/2\r_s )^4 =\f{4r^2}{K^2}.
\ee

For the above expressions to be well-defined, we require
that $1-b^2 >0$. Because $K$ is the increasing function of $r$
having its minimum at $r=2M$ the  sufficient condition is
$1-|2\pa_z C/M|^2>0$.
In (\ref{2.16}) we separated the factors $V$ which are most singular in the
near-horizon limit $r\rightarrow 2M$.

To show that the surface $r=2M$ is the infinite red-shift surface,
we solve the null geodesic equations in the metric (\ref{2.16}) in the limit
 $V\rightarrow 0$. Geodesic equations can be written either
in a form using the Christoffel symbols, or starting from the Lagrangian
corresponding to the metric \cite{chand}. 
Separating the leading in $V\rightarrow 0$ parts of the metric components, 
we obtain the Lagrangian in a form
\be
\l{g1}
{\cal L}=-\f{V {\dot{t}^2}}{2} + \f{{\dot{r}}^2 \b{g}_{rr}}{2V} + \f{\b{g}_{ra}\dot{z}_a
\dot{r}}{V^{1/2}} +\f{1}{2} \b{g}_{ab}\dot{z}_a\dot{z}_b
.\ee
Here dot is  derivative with respect to an affine parameter along
the geodesic $\tau$, bar over metric component means that the leading in
$V\rightarrow 0$ factor is written explicitly.
The Lagrange equations are
\ba
\l{g2}
 &{}&\f{d(tV)}{d\tau}=0,\\
\l{g3}
&{}& \f{\ddot{r}\b{g}_{rr}}{V} +\f{{\dot{r}}^2 \b{g}_{rr,r}}{V} + 
\f{{\dot{r}}^2 \b{g}_{rr}V_{,r}}
{2V^2} +\f{\b{g}_{ar}\dot{r}\dot{z}_a}{V^{1/2}} - \f{\b{g}_{ra}\ddot{z}_a}{V^{1/2}} +
\f{{\dot{t}}^2 V_{,r}}{2} +\f{1}{2} \b{g}_{ab,r}\dot{z}_a\dot{z}_b =0,\\
\l{g4}
&{}& -\f{\b{g}_{ar}{\dot{r}}^2 V_{,r} }{2V^{3/2}} + \f{\b{g}_{ar}{\dot{r}}^2 }{V^{1/2}}+
\f{ \b{g}_{ar}\ddot{r}}{V^{1/2}} +\b{g}_{ab}\ddot{z}_b +\b{g}_{ab,r}\dot{r}\dot{z}_b
+\f{1}{2}\b{g}_{bc,a}\dot{z}_b\dot{z}_c =0
.\ea   
We look for a solution in the asymptotic region $V\rightarrow 0$ in a form
\ba
\l{g5}
&{}&\dot{t}=\f{E}{V} ,\\
\l{g6}
&{}&\dot{r}=C+C_1 V^{1/2}+\cdots ,\\
\l{g7}
&{}& \dot{z}_a =A_a V^{-1/2}+A_{1a}+\cdots .
\ea
Substituting the Ansatz in Eqs. (\ref{g2})-(\ref{g3}) 
and separating the leading in $V\rightarrow 0$ terms, we have
\ba
\l{g8}
&{}&V^{-2} [E^2 -\b{g}_{rr} C^2  - \b{g}_{ra} A_a C ]=0,\\
\l{g9}
&{}& V^{-3/2} [\b{g}_{ra} C+ \b{g}_{ab} A_b ] =0
.\ea
Solving the system (\ref{g8})-(\ref{g9}),  we obtain
\ba
\l{g10}
&{}&
A_b = -(g^{-1})_{ba}\b{g}_{ar}, \\
\l{g11}
&{}&
\l{g12}
E^2 =[\b{g}_{rr}-\b{g}_{rb} (g^{-1})_{ba} \b{g}_{ar} ]C^2
.\ea
Because of  translation invariance  of the metric in $t$,
the system of equations
(\ref{g2})-(\ref{g4}) has the first integral ${\cal L}={\cal  H}=0$.
Substituting the  Ansatz  with the solution (\ref{g10})-(\ref{g11}),
we find that in the main order in
$V\rightarrow 0$ the relation ${\cal H}=0$ is satisfied identically.
From (\ref{g5}) and (\ref{g6}) in the main order in $V\rightarrow 0$ we obtain
\be\l{g13}
\f{dr}{dt}=\pm |E|V(r).
\ee
From this relation it follows that the surface $r=2M$ is the surface of infinite
redshift \cite {LL}.

\section{Diffeomorphisms preserving the near-horizon form of the metric}

In this section we study general diffeomorphisms which preserve the form of 
the metric in the near-horizon. To discuss the near-horizon geometry, we
consider the leading in $x=r-2$ terms in the metric. 
Requiring that the functional form of the leading terms is preserved under
the action of diffeomorphisms, we obtain the restrictions on the vector
fields generating diffeomorphisms. Transformations preserving the gauge 
conditions $g_{rt}=g_{at}=0$ result in time independence of generators 
of transformations, from transformations of other metric components 
follow restrictions on the $x$-dependence of generators.

Separating the leading 
in $x$ behavior in
the metric components, we present the metric (\ref{2.16}) as
\ba
\l{4.1}
&{}&ds^2 = {g}_{tt}dt^2 +{g}_{rr} dr^2 +{g}_{ra}dr dz^a +{g}_{ab}dz^a dz^b=
\\\nonumber
&{}&=(-g_{tt,1} x +O(x^2 ))dt^2 + \left(\f{g_{rr,-1}}{x} +O(x^{-1/2} )\right)dx^2+
2\left(\f{g_{ra,-1/2}}{x^{1/2}}+O(x^0 \right)dxdz^a + \\\nonumber
&{}&+(g_{ab,0}+O(x^{1/2})dz^a dz^b .
\ea
Under the diffeomorphisms generated by vector fields $\c^k $ the metric
components are transformed as
\be
\l{4.4}
L_\chi g_{mn} =\chi^k\pa_k {g}_{mn}+\pa_m \chi^k {g}_{kn} + \pa_n \chi^k  {g}_{mk}
\ee
We look for the components of the vector fields $\c^k$ in a form of expansion
in powers in $x^{1/2}$
$$
\c^k = \c^k_0 +x^{1/2}\c^k_{1/2} +x \c^k_{1} +\cdots
$$
The metric (\ref{2.1}) is written in the gauge
$$
{g}_{rt}={g}_{ta}=0
.$$
Transformations preserving the gauge conditions are
\ba
\l{4.6}
L_{\c}{g}_{rt}=\pa_r \c^t {g}_{tt} +\pa_t\c^r {g}_{rr}+\pa_t \c^a {g}_{ar}=0\\
\l{4.7}
L_{\c}{g}_{at}=\pa_a \c^t {g}_{tt}+\pa_t\c^r {g}_{ra}+\pa_t \c^b {g}_{ba}=0
.\ea
From the equations (\ref{4.6})- (\ref{4.7}) it follows
that
\ba
\l{4.81}
&{}&\dot{\c}^r_0 =\dot{\c}^r_{1/2} =\dot{\c}^r_{1}=0,\\
&{}&\dot{\c}^a_0 =\dot{\c}^a_{1/2}=0.
\ea
Here  "dot" denotes differentiation over $t$.

Transformations preserving the leading in $x$ behavior of other  metric 
components are
\ba
\l{4.10}
&{}& L_\c g_{tt} = \c^r\pa_r g_{tt}+2\pa_t \c^t g_{tt} =O(x),\\\nonumber
&{}& L_\c g_{rr}= \c^r\pa_r g_{rr}+\pa_a \c^a g_{rr} + 2\pa_r \c^a g_{ar}
+ 2\pa_r \c^r g_{rr} =O(x^{-1}),\\\nonumber
&{}& L_\c g_{ar}=  \c^r\pa_r g_{ar}+\c^b\pa_b g_{ar} + \pa_a \c^r g_{rr} +
\pa_a \c^b g_{br} + \pa_r \c^r g_{ar} + \pa_r \c^b g_{ab} =O(x^{-{1/2}}),\\\nonumber
&{}& L_\c g_{ab } =(\c^r\pa_r +\c^c\pa_c ) g_{ab}+\pa_a \c^r g_{rb} +
\pa_a \c^c g_{cb} +\pa_b \c^r g_{ra} + \pa_b \c^c g_{ca}=O(x^0 ).
\ea
From the transformation of the $g_{rr}$-component it follows that
\be
\l{4.1a}
\c_0^r =\c_{1/2}^r =0.
\ee
From transformations (\ref{4.10}) we extract transformations of
the leading-order terms of metric components
\ba
\nonumber
&{}&
\d_\c g_{tt}=\c_1^r g_{tt} +\c^a_0\pa_a g_{tt} +\dot\c^t_0 g_{tt}\\\nonumber
&{}&
\d_\c g_{rr,-1}=\c^r_1 \,g_{rr,-1}+\c^a_{1/2}\, g_{a r, -1/2} +\c^a_0\,
\pa_a g_{rr,-1}\\\nonumber
&{}&
\d_\c g_{a r, -1/2} =\f{1}{2}\,\c^r_1\,g_{a r, -1/2}+
\c_0^b\,\pa_b g_{a r, -1/2}+
\pa_a\c^b_0\, g_{b r, -1/2} +\f{1}{2}\c^b_{1/2}\,g_{ab,0}\\\nonumber
&{}&
\d_\c g_{ab ,0}=\c^c_0 \pa_c\,  g_{ab ,0} +
 \pa_a\c^c_0\, g_{cb ,0}+\pa_b\c^c_0\, g_{ca ,0}\\\nonumber
&{}&\d_\c g_{ab ,1/2}= \f{1}{2}\c^r_1\, g_{ab ,1/2} + \pa_a\c^r_1\,
g_{rb ,-1/2}+\pa_b\c^r_1\,g_{ra ,-1/2} +\c^c_0\pa_c\, g_{ab ,1/2} +
\c^c_{1/2}\pa_c\, g_{ab ,0}+\\
\l{4.11}
&{}& +\pa_a\c^c_0\,g_{cb ,1/2} + \pa_a\c^c_{1/2}\, g_{cb,0}
+\pa_b\c^c_0\,g_{ca ,1/2} + \pa_b\c^c_{1/2}\, g_{ca,0}
\ea
The leading-order parts of the metric components are transformed
through the functions 
\be
\l{4.13}
x\c_1^r (z,\z),\quad \c_0^z (z,\z),\quad \c_0^{\z} (z,\z),\quad 
x^{1/2}\c_{1/2}^z (z,\z),\quad x^{1/2}\c_{1/2}^{\z} (z,\z),\quad {\c}^t_0 (t,z.\z)
.\ee 
The variations of the metric components are  used in Sect.5
for calculation of the surface charge corresponding to asymptotic horizon
symmetries.
The Lie brackets of vector fields generating the near-horizon transformations 
are
\be
\l{4.12}
[\c_{(1)}, \,\c_{(2)}]^k = \c_{(12)}^k
,\ee
where
\ba\nonumber
&{}&\c_{(12),0}^t =\c^t_{(1),0}\stackrel{\leftrightarrow}{\pa}_t\c^t_{(2),0} +
\left(\c^b_{(1),0}\pa_b \c^t_{(2),0}-(1\leftrightarrow 2)\right),
\\\nonumber
&{}&\c^r_{(12),1}=
\c^b_{(1),0}\pa_b \c^r_{(2),1}-(1\leftrightarrow 2),
\\\nonumber
&{}& \c^a_{(12),0}=
\c^b_{(1),0}\stackrel{\leftrightarrow}{\pa}_b\c^a_{(2),0},\\
&{}& \c_{(12),1/2}^a=
\left(\c^b_{(1),0}\pa_b\c^a_{(2),1/2}-
(1\leftrightarrow 2)\right)
+1/2 \left(\c^r_{(1),1}\c^a_{(2),1/2}-
(1\rightarrow 2)\right).
\ea

\section{Supertranslations preserving the static\newline
gauge of the metric}

The metric (\ref{2.1}) was obtained in \cite{comp2} in the static gauge 
$\t{g}_{\r a}=
\t{g}_{t a}=0$.
The aim of this section is to find which supertranslations preserve both
the gauge and the form of the metric (\ref{2.16})  at the horizon.
We shortly review the supertranslatons preserving the gauge of (\ref{2.1}).
Next, we write the generator of supertranslations in the case of
the metric (\ref{2.16}). We find conditions on the generator of supertranslations
under which the functional form of the component 
$g_{tt}$ at the horizon is preserved.
It is shown that these conditions  are sufficient to preserve the 
near-horizon form of all the components of the metric.

Generator of supertranslations preserving the static gauge of (\ref{2.1})
 was found in \cite{comp2} and has the form
\be
\l{3.1}
\xi_T =T_{00}\pa_t -(T-T_{00})\pa_ \r +F^{ab}D_a T D_b
,\ee
where
$$
F^{ab}= \f{C^{ab}-2\g^{ab}(\r -E)}{2((\r -E)^2 -U)^2 }.
$$

Horizon of the metric (\ref{2.1}) is located at the surface  
\be
\l{3.4}
\r_H (z,\z )=C+\sqrt{1/4 -D_a C D^a C}
.\ee
Horizon of the metric (\ref{2.16}) is located at the surface $r=2$.

Generator of supertranslations preserving the static gauge of the metric (\ref{2.16})
is obtained from the generator (\ref{3.1}) by the coordinate transformation
\ba
\l{3.5}\nonumber
&{}&
\c^r_T =\xi^\r \f{\pa r}{\pa\r}+\xi^a \f{\pa r}{\pa z^a }=
\xi^\r \f{\pa r}{\pa\r_s}\f{\pa \r_s}{\pa \r } +
\xi^a \f{\pa r}{\pa\r_s}\f{\pa \r_s}{\pa z^a},\\
&{}&\c^a_T =\xi^\r \f{\pa z^a}{\pa\r}+\xi^b \f{\pa z^a}{\pa z^b}=\xi_T^a.
\ea
From (\ref{2.2}) and (\ref{2.5}), we have 
$$
\pa r/\pa\r_s =\f{K^2 -1}{K^2},\qquad \pa\r_s/\pa\r =\sqrt{1-b^2},\qquad
\pa\r_s/\pa z^a =\f{K}{4}[-2b_a\sqrt{1-b^2}+D_a b^2 ]
$$ 

In variables $t,r$ of the metric (\ref{2.16}) generator of supertranslations 
takes the form
\be
\l{3.6}
\c_T =\c_T^t\pa_t +\c_T^r\pa_r +\c_T^a \pa_a ,
\ee
where
\ba
\l{3.7}
&{}&
\c_T^t =T_{00},\\\nonumber
&{}&
\c_T^r =\f{K^2 -1}{K^2}\left(-(T-T_{00})\sqrt{1-b^2} +\f{K}{4}F^{ab}D_b T
(-2b_a\sqrt{1-b^2}+D_a b^2)\right),\\\nonumber
&{}&
\c_T^a = F^{ab}D_b T.
\ea
Here $F^{ab}(r,z^a )= F^{ab}(\r=\r(r,z^a ),z^a )$.
In the near-horizon region  $r=2+x,\,\,\, |x|\ll 1$, we have
\be
\l{3.9a}
K\simeq 1+\sqrt{2x},\,\quad b_a=b_{a 0} (1-\sqrt{2x}),\,\,\, b_{a 0} =2\pa_a C .
\ee
For $x\ll 1$ the component $g_{tt}$ is
\be
\l{3.9}
g_{tt}=\f{x}{2}+O(x^2 ).
\ee
Acting by the generator of supertranslations on the component $g_{tt}$, we obtain 
\be
\l{3.10}
L_{\c_T}g_{tt} = \f{2}{r^2}\f{K^2-1}{K^2}\left(-(T-T_{00})\sqrt{1-b^2} +\f{K}{4}
(-2b_a\sqrt{1-b^2}+D_a b^2)\right)
.\ee
In the near-horizon region $(K^2 -1 )/K^2 =O(x^{1/2} )$.
To have the  same form of the transformed component  $g_{tt}$ as in (\ref{3.9}),
 the second factor  in brackets in (\ref{3.10}) should be of order  $O(x^{1/2} )$
or less, i.e.
\be
\l{3.11a}
-(T-T_{00})\sqrt{1-b^2} +\f{K}{4}F^{ab}D_b T
(-2b_a\sqrt{1-b^2}+D_a b^2)=O(x^{1/2}).
\ee
 This imposes condition on $T(z,\z )$
\be
\l{3.11}
[-(T-T_{00})\sqrt{1-b^2} +\f{1}{4}F^{ab}D_b T
(-2b_{a}\sqrt{1-b^2}+D_a b^2)]_{r=0}=0
.\ee
Eq.(\ref{3.11}) for $T(z,\z )$  is solved in Appendix A.
From ondition (\ref{3.11a}) it follows that the generator of supertranslations in the 
near-horizon region 
has the following stracture
\be
\l{3.13}
\c_T =O(x^0 )\pa_t +O(x)\pa_x +O(x^0 ) \pa_a
.\ee
It is seen that the generator is of the form found in Sect. 3, and its 
action preserves
 the near-horizon behavior of all the metric components.

Transformations  (\ref{3.1}) and (\ref{3.6}) form a commutative 
algebra under the modified
bracket \cite{comp2}
\be
\l{3.12}
[\xi_1 ,\xi_2 ]_{mod} =[\xi_1 ,\xi_2 ] -\d_{T_1}\xi_2 +\d_{T_2}\xi_1 .
\ee

\section{Surface charge of asymptotic horizon symmetries}

In this section, first, we calculate the variation of the surface charge
 corresponding to the asymptotic horizon symmetries, and, next, discuss
 conditions on the form of the metric at the horizon which allow for 
integrating  the variation of the charge over the space of metrics 
to a closed form. We consider a particular example of such metrics.

We use the short notations of Sect. 3:
\ba
\l{5.1}
\begin{array}{c} {}\\g_{mn}=\\{}\\{}\end{array}
\left|\begin{array}{cccc}-\b{g}_{tt} x &0&0&0\\
0 & \b{g}_{rr}/x & \b{g}_{rz}/\sqrt{x} & \b{g}_{r\z}/\sqrt{x}\\
0 & \b{g}_{rz}/\sqrt{x} & \b{g}_{zz} & \b{g}_{z\z}\\
0 & \b{g}_{r\z}/\sqrt{x} & \b{g}_{z\z} & \b{g}_{\z\z}
\end{array}\right|
\qquad
\begin{array}{c} {}\\g^{mn}=\\{}\\{}\end{array}
\left|\begin{array}{cccc}-\b{g}^{tt}/ x &0&0&0\\
0 & \b{g}^{rr}x & \b{g}^{rz}\sqrt{x} & \b{g}^{r\z}\sqrt{x}\\
0 & \b{g}^{rz}\sqrt{x} & \b{g}^{zz} & \b{g}^{z\z}\\
0 & \b{g}^{r\z}\sqrt{x} & \b{g}^{z\z} & \b{g}^{\z\z}
\end{array}\right|
,\ea
where $\b{g}_{mn}= g_{mn,k}$ and $g_{mn,k}$  are the leading-order in $x$ parts
of metric components.
The elements of the corresponding  matrix  of metric
  variations calculated in
Sect. 3 are denoted as  $\d{g}_{mn} =h_{mn},\,\,\d\b{g}_{mn} =\b{h}_{mn}$.  
The inverse  variations are defined as
$h^{mn}={g}^{mk}h_{kl}{g}^{ln},$
and the trace of variations is $h =h_{mn}g^{mn}=\b{h}$.

Variation of the surface charge of the asymptotic symmetries is calculated
following \cite{bar_br}
\ba
\l{5.4}
\not\d_\c \hat{Q} (g,h) =\f{1}{4\pi}\int (d^2 x)_{ab}\sqrt{g^{(2)}}
\left[\c^a\nabla^b h -\c^a\nabla_\s h^{b\s} +\c_\s\nabla^a h^{b\s}+
\f{1}{2}h\nabla^a\c^b \right.\\\nonumber
+\left.\f{1}{2}h^{a\s}(\nabla^b \c_\s -\nabla_\s \c^b)-
 (r\leftrightarrow t) \right].
\ea
In present case
 $(d^2 x)_{ab}= (1/4) \e_{abmn}dx^m dx^n$ where $ m , n =z,\z  $ and $a,b =r,\, t$.
Integration is performed over the sphere of the radius  $r=2+x$, determinant
of the metric is
$ g^{(2)}=g_{zz}g_{\z\z}-g^2_{z\z}$, where ${g}^{(2)}=\b{g}^{(2)}/(2 +x)^2 $.
The limit $x=0$ is taken
 at the and of the calculation. 

Calculating contributions of the leading-order in $x\rightarrow 0$ 
parts of the five terms in the  
integrand of (\ref{5.4}), we obtain
\ba
\nonumber
&{}& 1.\quad\c^r\nabla^t h-\c^t\nabla^r h =\c^r g^{tt}\pa_t h -
\c^t g^{rr}\pa_r h - \c^t g^{ra}\pa_a h =O(x^{1/2}). \\\nonumber
&{}& 2.\quad -\c^r\nabla_s h^{ts} +
\c^t \nabla_s h^{rs}=- \c^r \nabla_t h^{tt}+
\c^t(\nabla_r h^{rr} +\nabla_a h^{ra}+ \nabla_t h^{rt})=       \\\nonumber
&{}&\left(-\b{g}^{rr}\f{\b{h}_{tt}}{\b{g}_{tt}} +\b{h}^{rr}\right)+O(x^{1/2}) .\\\nonumber
&{}& 3.\quad \c_s \nabla^r h^{ts}-\c_s \nabla^t h^{rs} =
 \c_t \nabla^r h^{tt} -\c_r\nabla^t  h^{rr}-\c_a\nabla^t  h^{ra}=\\\nonumber 
&{}&\f{\c^t_0}{2}\left(\b{g}^{rr}
\f{\b{h}_{tt}}{\b{g}_{tt}}-\b{h}^{rr}\right)+O(x^{1/2}).
\\\nonumber
&{}& 4.\quad \f{h}{2}(\nabla^r\c^t -\nabla^t\c^r )=
\f{h}{2}\left[(\b{g}^{rr}\nabla_r+\b{g}^{ra}\nabla_a )\c^t -\b{g}^{tt}
\nabla_t \c^t\right] = \f{\c^t_0}{2} \b{g}^{rr}h+ O(x^{1/2}).\\\nonumber
&{}& 5.\quad \f{1}{2}(h^{rs}\nabla^t\c_s -h^{ts}\nabla^r\c_s)=
\f{1}{2}(h^{rr}\nabla^t\c_r +h^{ra}\nabla^t\c_a-h^{tt}\nabla^r\c_t)=\\\nonumber
&{}&\f{\c^t_0}{4}\b{g}^{rr}\left[\f{\b{h}_{tt}}{\b{g}_{tt}}-
(\b{h}^{rr}\b{g}_{rr}\b{g}^{rr}+
\b{h}^{ra}\b{g}_{ar}\b{g}^{rr}+
\b{h}^{rr}\b{g}_{ra}\b{g}^{ar}+\b{h}^{ra}\b{g}_{ab}\b{g}^{br})\right]+ O(x^{1/2})=
\\\nonumber
&{}&=-\f{\c^t_0}{4}\left( \b{g}^{rr}\f{\b{h}_{tt}}{\b{g}_{tt}}+
\b{h}^{rr} \right)+O(x^{1/2}).
\\\l{5.5}
&{}& 6. \quad -\f{1}{2}(h^{rs}\nabla_s \c^t -h^{ts}\nabla_s \c^r )=
-\f{1}{2}(h^{rr}\nabla_r \c^t  +h^{ra}\nabla_a \c^t -h^{tt}\nabla_t \c^t )=\\
\nonumber
&{}&-\f{\c_0^t}{4}\left(\b{g}^{rr}\f{\b{h}_{tt}}{\b{g}_{tt}} 
+\b{h}^{rr}\right)+O(x^{1/2}).
\ea
In  item 2 we have used that
$$
\nabla_r h^{rr}=\pa_r h^{rr} +2\Gamma^r_{rr}h^{rr}+2\Gamma^r_{ra}h^{ra}=
\b{h}^{rr} -\b{h}^{rr} (\b{g}^{rr}\b{g}_{rr}+\b{g}^{ra}\b{g}_{ar}) +O(x^{1/2})=O(x^{1/2})
,$$
and in item 5 we used the identities
\ba
\nonumber
&{}&\b{g}^{rr}\b{g}_{rr}+\b{g}^{ra}\b{g}_{ar} =1,\\\nonumber
&{}&\b{g}^{rr}\b{g}_{ra}+\b{g}^{rb}\b{g}_{ba} =0.
\ea
Substituting  the expressions (\ref{5.5}) in (\ref{5.4}), we obtain
\be
\l{5.7}
\not\d_\c Q (g,h)= \f{1}{16\pi}\int dz d\z \, \sqrt{\b{g}^{(2)}}\f{\c^t_0}{2}\left(
\b{g}^{rr}h -\b{h}^{rr}-\b{g}^{rr}\f{ \b{h}_{tt}}{\b{g}_{tt}}\right).
\ee
The combination in the integrand can be presented as
\be
\l{5.9}
\b{g}^{rr}\b{h} -\b{h}^{rr}-\b{g}^{rr}\f{ \b{h}_{tt}}{\b{g}_{tt}}
=(\b{g}^{rr}\b{g}^{ab} -\b{g}^{ra}\b{g}^{rb})\b{h}_{ab}.
\ee
Denoting determinant of the 3D part of the metric
as $ \b{g}^{(3)}$, and using the identity
$$
\b{g}^{rr}\b{g}^{ab} -\b{g}^{ra}\b{g}^{rb}=\b{g}_{ab}\b{g}^{(3)}
,$$
we can write (\ref{5.9}) as
\be
\l{5.10}
(\b{g}^{rr}\b{g}^{ab} -\b{g}^{ra}\b{g}^{rb})\b{h}_{ab}=
(\b{g}_{zz}\d \b{g}_{\z\z} +\b{g}_{\z\z}\d \b{g}_{zz}-
2\b{g}_{z\z}\d \b{g}_{z\z} )/\b{g}^{(3)}=\f{\d \, \b{g}^{(2)}}{ \b{g}^{(3)}}
\ee
A sufficient condition of integrability of (\ref{5.7}) is $ \b{g}^{(3)}=
f( \b{g}^{(2)} )$, where $f$ is an integrable function. 
This case is realized, if supertranslation field depends 
on $|z^2|$, or, in spherical
 coordinates $\th ,\p $, only on $\th$, i.e. $C(z,\z ) =C(|z|^2 )= \t{C}(\th )$.
In this case, because of the identity $\b{g}_{rr}\b{g}_{\th\th} -{\b{g}_{r\th}}^2=
\b{g}_{\th\th}$ (see Appendix B), we have $ \b{g}^{(3)}=
(\b{g}_{rr}\b{g}_{\th\th} -{\b{g}_{r\th}}^2 )\b{g}_{\p\p} =\b{g}^{(2)} $.  
 The charge is
\be
\l{5.11}
\hat{Q} = \f{1}{16\pi}\int {\c_0^t}(\b{g}_{\th\th}\b{g}_{\p\p})^{1/2}d\th d\p \sin\th 
+\hat{Q}_0.
\ee
With $\c_0^t$ independent of $\th$ the charge is proportional 
to the surface of the horizon,
i.e. to the entropy of the black hole.

\section{Conclusions}

In this paper we studied the near-horizon symmetries
of the metric of a black hole containing
  supertranslation field.
To study general transformations preserving the near-horizon form of
the metric, we transformed the metric to a coordinate system in which the
horizon of the metric is located at the surface $r=2M$.
Solving the geodesic equations for null geodesics, it was shown
that the surface $r=2M$ is the surface of infinite redshift.
We reviewed the action of the generators of supertranslations 
preserving the gauge of the metric
and found a class of supertranslations which also preserve the
near-horizon form of the metric.

Next, we determined the form of generators of asymptotic horizon symmetries 
preserving the  near-horizon form of  the metric with supertranslation field.
Variations of the metric components under the action of generators of
near-horizon symmetries were obtained. Using the 
 variations of the metric components, we obtained variation 
of the surface charge corresponding to the asymptotic horizon symmetries.

Studying the form of the variation of the surface charge, we found
a sufficient condition of integrability of the variation of the surface charge 
over the space of metrics to a closed expression. 
In the case of metrics with the supertranslation field depending on $|z^2|$, or,
in spherical variables $\th , \p$,
 with dependence only on $\th$, it was shown that
due to specific relations between the metric components,
it is  possible
to integrate the variation of the surface charge and obtain the charge of
horizon symmetries in a closed form.  The charge is proportional to the 
horizon surface and can be interpreted as entropy of the black hole.

Horizon symmetries of a class of metrics with the near-horizon form
\be
\l{1.7}
ds^2 = -2\k \r dv^2 +2dv\,d\r +2\r h_a ( x) dv\,dx^a +
(\o_{ab}(x)+\r \la_{ab}(x))dx^a dx^b,\qquad x^a =z,\,\z
\ee
were previously studied in many papers, as examples, in
\cite{carlip,koga,hotta,donnay41,donnay4,akh,set}.
 Horizon of the metric is located at $\r =0$, and $\r$ is a distance of the
surface $\r =const$ from the horizon.
The metric is written in the gauge $g_{\r\r}=g_{\r a}=0$.

In the near-horizon region the metric components are
expanded in power series in $\r$ \cite{booth}.
In contrast to this case, the metrics considered in the present paper
 are expanded in series of $(r-2M)^{1/2}$.
In the limit of vanishing
 supertranslation field the terms with fractional powers of $r-2M$
vanish.

The charge of the asymptotic horizon symmetries for metrics (\ref{1.7}),
in the case $\k =const$
was obtained in \cite{donnay41,donnay4} in a form
$$
Q=\f{1}{16\pi G}\int dz d\z \sqrt{\g}(2 T\k\Omega-y^a\theta_a\Omega )
.$$
Here $T(z,\z )$ is a part of the asymptotic Killing vector $\c^v (z,\z ,v)$, 
and the metric
component $\o_{ab}$ has diagonal form $\o_{ab} =\g_{ab}\Omega $.
The volume $\sqrt{\g}\,\Omega\, dz d\z$ is an analog of the volume
$\sqrt{ \b{g}_{\th\th}\b{g}_{\p\p}}\,\sin\th d\th d\p$ in (\ref{5.11}),
and $T$ is an analog of $ \c^t$. It is seen that the corresponding 
structures of the charges in both cases
are similar.

The  near-horizon transformations for the pure Schwarzschild  metric
were considered in
\cite{koga} and refs. therein. In the limit of vanishing supertranslation field
the results of these papers and the present paper are identical.


\section{Appendix A}
\setcounter{equation}{0}
\renewcommand{\theequation}{A\arabic{equation}}

In this Appendix we find general solution of Eq. (\ref{3.11})
\be
[-T\sqrt{1-b^2} +\f{1}{4}F^{ab}D_b T
(-2b_{a}\sqrt{1-b^2}+D_a b^2)]_{r=2}=0
\ee
which can be presented in a form 
\be
\l{b2}
T + F^a D_a T=0
,\ee 
where 
$$
F^a = \f{1}{2}F^{ac}(b_c + \pa_c \sqrt{1-b^2})|_{r=2} 
$$
Following the general rules of
solving the differential equations with partial derivatives \cite{step},
we consider  a function
$W(T,z, \z )$ satisfying the equation
\be
\l{b3}
T\f{\pa W}{\pa T} + F^z \f{\pa W}{\pa z} + F^{\z} \f{\pa W}{\pa \z}=0
.\ee
 Eq.(\ref{b3}) is solved by writing  the system of ordinary differential equations
\be
\l{b4}
\f{d T}{T} =\f{d z}{F^z}= \f{d \z}{F^{\z} }
.\ee
Let the independent first integrals of the Eq. (\ref{b4}) be
\ba
\l{b5}
\psi_1 (T,z,\z ) =C_1, \qquad \psi_2 (T,z,\z ) =C_2.
\ea
The general solution of the Eq. (\ref{b3}) for  $W(T, z,\z )$ is
\be
W=f(\psi_1 ,\psi_2 ),
 \ee
where $f$ is an arbitrary smooth function.
The function $T(z,\z )$ is determined from the equation
\be
f(\psi_1 ,\psi_2 )=0.
\ee


\section{Appendix B}
\setcounter{equation}{0}
\renewcommand{\theequation}{B\arabic{equation}}

If supertranslation field depends $C(z, \z )$ depends only on $|z|^2$, 
 it is convenient to parametrise the unit sphere by $\th ,\p$
coordinates. In these coordinates the metric  (\ref{2.16}) takes a form
\ba\nonumber
&{}&ds^2 =g_{\m\n} dx^\m dx^\n
=-V dt^2 +\f{dr^2 \b{g}_{rr}}{V} +\f{2drd\th \b{g}_{r\th}}{V^{1/2}}+
d\th^2 \b{g}_{\th\th}+ d\p^2 \sin^2\th\b{g}_{\p\p}\\\nonumber
&{}& =-V dt^2 +  \f{dr^2}{V(1-b^2 )} +2drd\th\f{br (\sqrt{1-b^2}-b' )}{(1-b^2 )V^{1/2}}
+\\\l{a1}
&{}& + d\th^2 r^2\f{ (\sqrt{1-b^2}-b' )^2}{(1-b^2 )} +d\p^2 r^2\sin^2\th (b\cot\th
-\sqrt{1-b^2 })^2
,\ea
where $b =2\pa_\th C(\th )/K$. 
It is explicitly verified that  the metric components are connected by a relation
\be
\l{a2}
{g}_{rr}{g}_{\th\th}- {g}_{r\th}^2 =\f{{g}_{\th\th}}{V}.
\ee

\vspace*{1cm}

{\large\bf Acknowledgments}

This work was partially supported by the Ministry of Science and Higher Education of
Russian Federation under the project 01201255504.


\end{document}